\documentclass[conference,a4paper]{APSIPA2021}
\usepackage{multirow}
\usepackage{graphicx}
\usepackage{amsmath}
\usepackage[psamsfonts]{amssymb}
\usepackage{amsxtra}
\usepackage{threeparttable}
\usepackage{cite}
\usepackage{stfloats}
\usepackage[justification=centering]{caption}
\usepackage{subfigure}

\usepackage{geometry}
\usepackage{fancyhdr}

\begin{document}

\title{Acoustic Pornography Recognition Using Convolutional Neural Networks and Bag of Refinements}

\author{%
\authorblockN{%
Lifeng zhou\authorrefmark{1}, Kaifeng Wei\authorrefmark{1}, Yuke Li\authorrefmark{1}, Yiya Hao\authorrefmark{2}, Weiqiang Yang\authorrefmark{1} and Haoqi Zhu\authorrefmark{1}
}
\authorblockA{%
\authorrefmark{1}
Netease YiDun  AI Lab, Hangzhou, China}
\authorblockA{%
\authorrefmark{2}
Netease CommsEase  AudioLab, Hangzhou, China \\
E-mail: \{hzzhoulifeng,hzweikaifeng,liyuke,haoyiya,yangweiqang,zhuhaoqi\}@corp.netease.com}
}

\maketitle

\begin{abstract}
 A large number of pornographic audios publicly available on the Internet seriously threaten the mental and physical health of children, but these audios are rarely detected and filtered. In this paper, we firstly propose a convolutional neural networks (CNN) based model for acoustic pornography recognition. Then, we research a collection of refinements and verify their effectiveness through ablation studies. Finally, we stack all refinements together to verify whether they can further improve the accuracy of the model. Experimental results on our newly-collected large dataset consisting of 224127 pornographic audios and 274206 normal samples demonstrate the effectiveness of our proposed model and these refinements. Specifically, the proposed model achieves an accuracy of 92.46\% and the accuracy is further improved to 97.19\% when all refinements are combined. 
\end{abstract}

\section{Introduction}
With the development of the Internet, substantial pornographic audios are uploaded freely every day, which raises the necessity of filtering  them. However, few works have been proposed to detect pornographic audios as far as we know. Banaeeyan et al. \cite{banaeeyan2019acoustic} proposed to use pith and Mel-Frequency Cepstrum Coefficients (MFCCs) \cite{ittichaichareon2012speech} as inputs to the KNN classification model to classify pornographic audios. Wazir et al. \cite{wazir2019acoustic} proposed to use recurrent neural network \cite{zeineldeen2020layer, zeyer2019comparison} to detect pornographic audios and achieved an accuracy of 86.50\% on the dataset consisting of 1000 audio samples. However, convolutional neural networks which are the dominating architecture for classification are rarely applied to acoustic pornography recognition.

Since the introduction of AlexNet  \cite{krizhevsky2012imagenet} in 2012, various new architectures have been proposed, such as VGG \cite{simonyan2014very}, Inception \cite{szegedy2015going}, DenseNet \cite{huang2017densely} and ResNet \cite{he2016deep}. With the development of model architecture, a lot of refinements have been proposed to improve the accuracy of convolutional neural networks. Attention mechanisms such as 
Squeeze-and-Excitation (SE) attention \cite{hu2018squeeze}, Convolutional Block Attention Module (CBAM) \cite{woo2018cbam} and coordinate attention (CA) can be used to guide the model to focus on important patterns and enhance feature learning. Pooling methods \cite{kao2020comparison} such as average pooling, max pooling and attentive pooling \cite{santos2016attentive} can reduce the spatial redundancy information and increase the invariance of translation transformation of neural networks. Label smoothing \cite{szegedy2016rethinking, yao2019robust} is a very simple regularization mechanism that avoids over-fitting by introducing  smoother ground-truth labels. Knowledge distillation \cite{hinton2015distilling, jung2020knowledge} transfers knowledge from a big model to a smaller one, which can bring consistent improvement without increasing model complexity. Warmup \cite{he2016deep, goyal2017accurate} uses a small learning rate at the start of training, then goes back to the original learning rate schedule to train the model to convergence.

Although convolutional neural networks and these refinements have achieved great success in image classification and some speech tasks, their performance on acoustic pornography recognition is still less researched. In this paper, we propose a convolutional neural network based model for acoustic pornography recognition  and examine how much the accuracy can be improved by a collection of refinements on our newly-collected large dataset.

Our main contributions can be summarized as follows:
\begin{enumerate}
\item To the best of our knowledge, this is the first time that a convolutional neural network based model is used for acoustic pornography recognition. 
\item We research a collection of general refinements which have been proved effective in classification and targeted refinements for acoustic pornography recognition. 
\item 
We build a large dataset including 224,127 pornographic audios and 274,206 normal samples and conduct extensive experiments on this dataset to demonstrate the effectiveness of the model and these refinements.
\end{enumerate}

\begin{figure}[h]
  \centering
  \includegraphics[width=\linewidth]{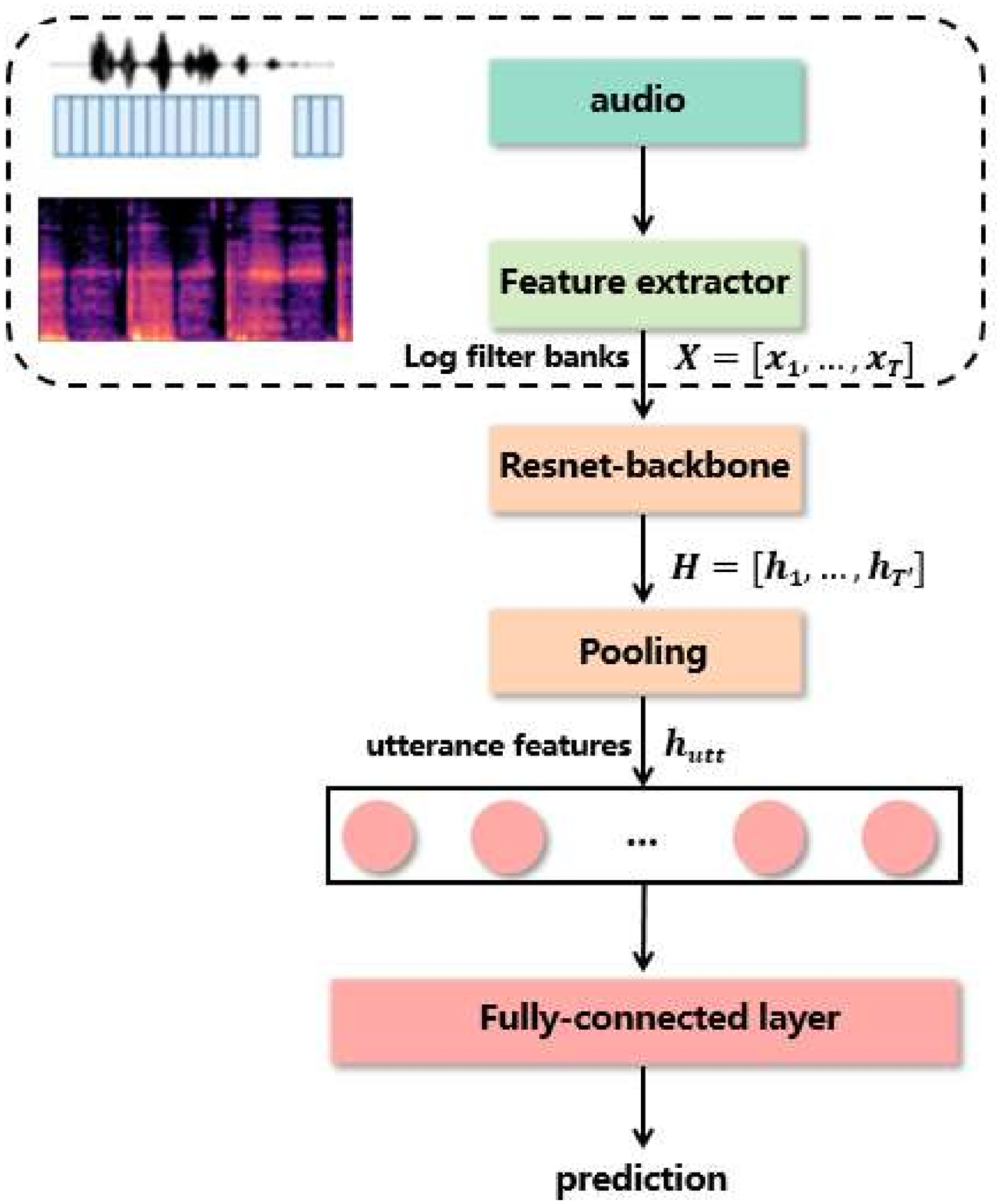}
  \caption{The pipeline for acoustic pornography recognition.}
  \label{fig:architecture}
\end{figure}

\section{Methods}
In this section, we firstly introduce the pipeline for acoustic pornography recognition. Then two targeted refinements for acoustic pornography recognition are delivered. Lastly, details of a series of general refinements for classification are shown one by one. 

\subsection{Pipeline }
The pipeline for acoustic  pornography recognition is shown as Figure \ref{fig:architecture}.  Firstly, we extract 80-dimensional log filter banks \cite{baskar2021eat}\cite{han2021time} $X=[x_1,...,x_T]$ from the audio, where $x_i \in R^{80}$ and $T$ is the number of frames. The extracted log filter banks are then fed into a convolutional neural network to extract 
frame level features $H= [h_1, ..., h_{T^{'}} ] \in R^{d \times T^{'}}$. Then the utterance level feature $h_{utt}$ is  obtained from $H$ with a pooling layer. Lastly the utterance level feature $h_{utt}$ is fed into a fully-connected layer to determine whether the audio is pornographic.

\subsection{Attention mechanisms}
Attention mechanisms have achieved great success in acoustic scene classification \cite{zhang2018multi}\cite{ren2018attention}, speaker verification \cite{desplanques2020ecapa}\cite{rahman2018attention}, and automatic speech recognition \cite{povey2018time}\cite{miao2019online}.
Considering the fact that different times and frequencies of the input $X$ have different influences on recognition results, we introduce attention mechanisms to help model focus on relevant time-frequency patterns.
In this section, we show some classic attention mechanisms including Squeeze-and-
Excitation (SE) \cite{hu2018squeeze}, Convolutional Block Attention
Module (CBAM) \cite{woo2018cbam} and Coordinate Attention (CA)  \cite{hou2021coordinate}.
\subsubsection{Squeeze-and-
Excitation}
Squeeze-and-Excitation attention \cite{hu2018squeeze} includes squeeze and excitation, where squeeze operation is used to squeeze global spatial information into a channel descriptor and excitation operation is used for channelwise feature adaptive recalibration.

 The squeeze function represented by \(F_{sq}\) takes $X \in \mathbb{R}^{C \times H \times W}$ as input, the \emph{c}-th channel of output \(\textbf z \in \mathbb{R}^C\) can be calculated as follows:
\begin{equation}
  z_c\ = F_{sq}(X_c) = \frac {1} {W \times H} \sum_{i=1}^{W} \sum_{j=1}^{H} X_c(i, j).
  \label{eq1}
\end{equation}

Next is the excitation operation, which is represented as:
\begin{equation}
  \hat X \ = X \bullet \textbf{s},
  \label{eq2}
\end{equation}
where \(\bullet\) means channel-wise multiplication. The formula for \(\textbf s\) is as follows:

\begin{equation}
  \textbf s\ = F_{ex}(\textbf z) = \sigma(W_2(\delta(W_1(\textbf z))),
  \label{eq3}
\end{equation}
where $W_1 \in R^{\frac{C}{r}\times C}$ and $W_2 \in R^{C\times\frac{C}{r}}$, \(\delta\) and \(\sigma\) represent ReLU \cite{nair2010rectified} and sigmoid function respectively.

\subsubsection{Convolutional Block Attention
Module}
Convolutional Block Attention
Module \cite{woo2018cbam} performs the attention on both channel and spatial dimension of the input features. Assuming the input is \(\textbf X \in \mathbb{R}^{C \times H \times W}\), then the formula for CBAM is as follows:

\begin{equation}
  \hat X\ =  F_s(F_c(X)) \otimes X,
  \label{eq4}
\end{equation}
where \(F_c\) and \(F_s\) represent  attention calculation of channel and spatial dimension respectively and  $\otimes$ denotes element-wise multiplication. The calculation formulas are shown as follows:
\begin{equation}
  F_c(X)\ = \sigma(W_{1}(W_{0}(X_{avg}^c)+W_{1}(W_{0}(X_{max}^c))),
  \label{f_c}
\end{equation}

\begin{equation}
  F_s(X)\ = \sigma(Con_{7\times7}([X_{avg}^s,X_{max}^s])),
  \label{f_s}
\end{equation}
where  $W_0 \in R^{\frac{C}{r}\times C}$, $W_1 \in R^{C\times\frac{C}{r}}$, \(\sigma\) represents the sigmoid function, [·,·] means concatenation and $Con_{7\times7}$ represents a convolution operation with the filter size of $7\times7$. 
$X_{avg}^c$ and $X_{max}^c$ denote average pooling and max pooling on channel dimension respectively. $X_{avg}^s$ and $X_{max}^s$ denote average pooling and max pooling on spatial dimension respectively.

\subsubsection{Coordinate Attention}
Coordinate attention \cite{hou2021coordinate} splits the spatial dimension into two directions and integrates them with the channel dimension to calculate the attention. 

Given an input \(\textbf X \in \mathbb{R}^{C \times H \times W}\), average pooling is computed along the vertical and horizontal directions as follows:

\begin{equation}
  z_c^h\ =  \frac {1} {W} \sum_{i=1}^{W} x_c^h(i),  z_c^w\ =  \frac {1} {H} \sum_{j=1}^{H} x_c^w(j).
  \label{}
\end{equation}

Then fuse the features of  two directions to get \(\textbf u\):

\begin{equation}
  \textbf u\ = F_1(Con_{1\times1}([\textbf z^h, \textbf z^w])),
  \label{}
\end{equation}
where [·,·] means concatenation operation, \(F_1\) means non-linear layers. Then \(\textbf s^h\) and \(\textbf s^w\) can be formulated as follows : 

\begin{equation}
  \textbf s^h = \sigma(Con_{1\times1}(\textbf u^h)),\textbf s^w = \sigma(Con_{1\times1}(\textbf u^w)).
  \label{}
\end{equation}

Finally, the output of coordinate attention is:
\begin{equation}
  \hat x_c(i, j)\ = x_c(i, j) \times s_c^h(i) \times s_c^w(j).
  \label{}
\end{equation}

\subsection{Pooling methods}
Since different frames of an audio have different weights on the recognition result, we propose different pooling methods which are used to aggregate frame-level feature $H$ to utterance-level feature $h_{utt}$. We
list the definition of three different pooling methods in Table ~\ref{tab:pooling methods},

\begin{table}[th]
  \caption{Pooling methods}
  \label{tab:pooling methods}
  \centering
  \begin{tabular}{ r@{}l  l }
     \hline
    \multicolumn{2}{l}{\textbf{Pooling methods}} &  
    \multicolumn{1}{l}{\textbf{Function}} \\
    \hline\hline
     & Average pooling &  $h_{utt}=\frac{1}{T^{'}}$$\sum_{t=1}^{T^{'}}$$ h_{t}$ \\
     & Attentive pooling  &  $h_{utt}=\sum_{t=1}^{T^{'}}$$a_t$$h_{t}$             \\
     & Max pooling  & 
      $h_{utt}[n]=\max\limits_{1<t<{T^{'}}}$$h_{t}[n]$
   \\
    \hline
  \end{tabular}
  
\end{table}
where $a_{t}$ is a learnable weighting factor, $1\le n \le d$ and $d$ is the dimension of $h_{t}$.
\subsection{Label smoothing}
Label smoothing  was first proposed by \cite{szegedy2016rethinking} to alleviate the over-fitting phenomenon of model training.

Assuming that \(z\) is the output of fully connected layer of the network, we usually use the following softmax function to get the probability for class \(i\):

\begin{equation}
  p_i = \frac{exp(z_i)}
  {\sum_{j=1}^{K}exp(z_j)},
  \label{}
\end{equation}
where $K$ is the number of classes. Therefore, the cross entropy loss \(l\) is:

\begin{equation}
  l = -\sum_{i=1}^{K}q_i log p_i,
  q_i=\left\{
    \begin{aligned}
      1, if(i=y), & \\
      0, if (i \neq y), &
    \end{aligned}
\right.
  \label{}
\end{equation}
where \(q_i\) represents ground truth label.
The above labels can easily lead to over-fitting. Therefore, label smoothing performs the following smoothing operations on the labels:
    
\begin{equation}
q_i=\left\{
    \begin{aligned}
      1- \epsilon, if(i=y), & \\
      \frac {\epsilon} {K-1} , if (i \neq y), &
    \end{aligned}
\right.
  \label{}
\end{equation}
where \(\epsilon\) is a smoothing parameter.

\subsection{Warmup}
In the training of deep neural networks, the setting of learning rate is important. The learning rate decay method commonly used in the past requires a large learning rate at the start of training. However, since the network parameters are relatively random at this time, a large learning rate has a greater impact on the early stage.  
\cite{he2016deep} proposed constant warmup, which used a constant small learning rate at the start of training, and then changed to the original large learning rate for regular decay after a period of training epoch. \cite{goyal2017accurate} adopted a new warmup strategy called gradual warmup, where the learning rate gradually increased from small to large throughout the warmup stage. After the warmup stage, it went back to the original learning rate schedule.

\subsection{Knowledge distillation}
Knowledge distillation \cite{hinton2015distilling} improves the ability of the student model by transferring the knowledge from the teacher model.

Assuming \(z_t\) and \(z_s\) are the last fully connected layer outputs of the teacher model and student model respectively, and \(g\) represents the ground truth label. The traditional cross-entropy-based classification loss function is as follows:

\begin{equation}
  L_{hard}\ = H(g, softmax(z_s)).
\end{equation}

The knowledge distillation loss is as follows:

\begin{equation}
  L_{KD}\ = T^{2}KL(softmax(z_t/T), softmax(z_s/T)),
\end{equation}
where $T$ is the temperature. The final training loss function of the student network is as follows:

\begin{equation}
  L_{all}\ = \lambda L_{hard} + (1-\lambda) L_{KD},
\end{equation}
where \(\lambda\) is a hyper-parameter to balance the two loss functions.

\section{Experiments Setup}

\subsection{Dataset}
Since there are no public datasets available for the acoustic pornography recognition task, we have checked 502387 videos on the Internet and extracted the soundtracks to collect a large-scale audio dataset from them. The newly assembled audio dataset consists of 224127 pornographic audios and 274206 normal samples which are all tagged by human. All the audios are resampled to 16KHZ with single channel by FFmpeg and the mean duration of these audios is ten seconds. In the future, we will open source the dataset. In all our experiments, the dataset is split into three partitions: training set, validation set and test set. Specifically, 70\% of the data are used for training, while 10\% of the data are used for validation and the rest are used for test. All results are measured by accuracy and evaluated on the test set.

\subsection{Implementation details}
For training, we obtain acoustic features of 80-dimensional log filter banks from training audios. All features are acquired from a 20ms window with a 10ms stride between frames. All experiments are conducted on the 8 NVIDIA GeForce GTX 1080Ti GPUs with batch-size of 1024. Adam optimizer is used for all experiments. All models are trained from scratch and the total number of epochs is 20. Except warmup experiments, all experiments' learning rate (lr) starts from 0.1 and is divided by 10 every 5 epochs. Constant warmup starts with lr = 1e-5 for 5 epochs, while gradual warmup starts with lr = 1e-5 and is linearly increased to lr = 0.1 over 5 epochs. After the warmup stage, we go back to the original learning rate decay schedule. We use ResNet18 and adopt average pooling as our baseline model and the training schedule is shown as Figure \ref{fig:architecture}. In knowledge distillation, we choose ResNet50 as the teacher model.

\section{Results}
\noindent \textbf{Baseline.} 
The baseline model achieves  92.46\% accuracy, which demonstrates the effectiveness of convolutional neural networks for the  task of acoustic pornography recognition.

\noindent \textbf{Attention.}  
As shown in Table ~\ref{tab:attention}, the accuracies of the three attention mechanisms are all higher than that of the baseline model, which illustrates the advantage of the attention mechanisms for acoustic pornography recognition. This is because attention mechanisms encourage model to concentrate more on classification-related patterns. What's more, different attention mechanisms have different effects on the performance. CBAM-ResNet18 is superior to SE-ResNet18 because SE-ResNet18 performs only channel attention, while CBAM-ResNet18 contains additional spatial attention, which helps the model focus on relevant time-frequency patterns. The reason CA-ResNet18 achieves higher accuracy than CBAM-ResNet18 is that CA encodes global spatial information to help the model accurately locate relevant time-frequency patterns, while CBAM only encodes local spatial information through a $7\times7$ convolutional layer. The characteristic of CA to capture long-range dependencies is more powerful to detect those pornographic audios whose frames of pornographic events are discontinuous and far apart.

\begin{table}[t]
  \caption{Comparisons of different attention methods.}
  \label{tab:attention}
  \centering
  \begin{tabular}{c c}
    \hline
    \textbf{Attention}            & \textbf{Acc}                  \\
    \hline\hline
    ResNet18                      & 92.46                          \\
    SE-ResNet18                   & 94.36                          \\
    CBAM-ResNet18                 & 94.93                          \\
    CA-ResNet18                   & \textbf{95.63}                 \\
    \hline
  \end{tabular}
\end{table}
\noindent \textbf{Pooling methods.} The results shown in Table ~\ref{tab:pooling} demonstrate that max pooling is the best pooling methods and both max pooling and attentive pooling are better than average pooling. This is because although these three pooling methods have similar performance on detecting long pornograhhic audios, max pooling and attentive pooling are superior to average pooling detecting short pornographic audios whose pornographic events duration is very short compared to the utterance duration. In this case, max pooling works best since once it detects the pornographic events at certain frames, these strong responses will be kept.
Although the weights of attentive pooling to pornographic frames are high, the ultimate responses are suppressed by the low weights of the non-pornographic frames.
The average pooling simply assigns the same weights to all frames which is not suitable for detecting short pornographic audios. 

\begin{table}[t]
  \caption{Comparisons of different pooling methods.  The default
baseline adopts average pooling.}
  \label{tab:pooling}
  \centering
  \begin{tabular}{c c}
    \hline
    \textbf{Pooling methods}      & \textbf{Acc}                \\
    \hline\hline
    average pooling             & 92.46                     \\
    attentive pooling        & 93.07                      \\
    max pooling               & \textbf{93.64}              \\
    \hline
  \end{tabular}
\end{table}

\noindent \textbf{Label smoothing.} From Table ~\ref{tab:lm} we can see that using label smoothing achieves higher accuracy than baseline. We believe the reason is that original ground-truth distribution for cross-entropy loss encourages the model to be more confident about its predictions and it may result in over-fitting. Label smoothing introduces smoother ground-truth distribution, which encourages the largest logit to be less different with all others. Intuitively, it is helpful for model regularization and avoiding over-fitting. As shown in Table ~\ref{tab:lm}, we also find the choice of hyper-parameter \(\epsilon\) has little effect on the results.

\begin{table}[t]
  \caption{Ablation studies of label smoothing. \(w/o\) means without and \(w/\) means with. \(\epsilon\)=0 is the same as not using label smoothing.}
  \label{tab:lm}
  \centering
  \begin{tabular}{c c c}
    \hline
    \textbf{Label smoothing}  &  \textbf{\(\epsilon\)} & \textbf{Acc}  \\
    \hline\hline
    w/o label smoothing   &  0   & 92.46                      \\
    w/ label smoothing   &  0.05   & 92.82                      \\
    w/ label smoothing   &  0.1    & \textbf{92.98}              \\
    w/ label smoothing   &  0.2    & 92.92              \\
    \hline
  \end{tabular}
\end{table}

\noindent \textbf{Warmup.}
As shown in Table ~\ref{tab:warmup}, the constant warmup and gradual warmup can achieve accuracies of 92.91\% and 93.26\% respectively, which indicates that using warmup strategy can lead to a consistent performance gain.
Compared with no warmup, constant warmup solves the optimization difficulties in the early stages of training and achieves higher accuracy (92.91\% vs 92.46\%). Especially, gradual warmup uses a progressive scheme to avoid training instability caused by a sudden increase of the learning rate, which further improves the accuracy to 93.26\%.
    
\begin{table}[t]
  \caption{Comparisons of different warmup methods.}
  \label{tab:warmup}
  \centering
  \begin{tabular}{c c}
    \hline
    \textbf{Warmup methods}      & \textbf{Acc}                \\
    \hline\hline
    no warmup                & 92.46                     \\
    constant warmup           & 92.91                          \\
    gradual warmup           & \textbf{93.26}                      \\
    \hline
  \end{tabular}
\end{table}

\noindent \textbf{Knowledge distillation.} Table ~\ref{tab:KD} shows the accuracies achieved by different temperature and \(\lambda\), which are two important hyper-parameters of knowledge distillation. First of all, with appropriate hyper-parameters, knowledge distillation can lead to considerable improvement in accuracy (93.66\% vs 92.46\%).
This is because knowledge distillation transfers the knowledge of a large model to a small one, which provides more effective supervision information for the training of the small model. Secondly, we find that as the temperature increases, the accuracy first rises and then falls. The reason is that model focuses on matching logits that are much more negative than the average at high temperature. This is potentially advantageous because the negative logits may convey useful information, but can also introduce more noise. Thus the choice of temperature is a trade-off between knowledge transfer and noise suppression. Thirdly, we find that the value of \(\lambda\) has little effect on the accuracy, and in our experiments, the setting of \(T\)=10 and \(\lambda\)=0.5 leads to the highest accuracy.

\begin{table}[t]
  \caption{Comparisons of different hyper-parameters of knowledge distillation. The accuracy of not using knowledge distillation is 92.46\%.}
  \label{tab:KD}
  \centering
  \begin{tabular}{c c c}
    \hline
    \textbf{Temperature}   & \textbf{\(\lambda\)}   & \textbf{Acc}    \\
    \hline\hline
     1           & 0.2 / 0.5 / 0.8         & 93.21 / 93.15 / 93.21   \\
    10           & 0.2 / 0.5 / 0.8         &  93.58 / \textbf{93.66} / 93.54  \\
    100         & 0.2 / 0.5 / 0.8          &  92.39 / 92.34 / 92.27 \\
    \hline
  \end{tabular}
\end{table}

\noindent \textbf{Refinements combining.}
 The accuracies for applying these refinements alone and one-by-one are
shown in Table ~\ref{tab:all}. By stacking coordinate attention (CA), max pooling (MP), label smoothing (LS), gradual warmup (GW) and knowledge distillation (KD), we have steadily improved the accuracy from 92.46\% to 97.19\%. 

\begin{table}[h]
  \caption{Performances of combining all refinements. LS adopts \(\epsilon\)=$0.1$, KD adopts T=$10$ and \(\lambda\)=$0.5$.}
  \label{tab:all}
  \centering
  \begin{tabular}{c c c c c c}
    \hline
    \textbf{CA} & \textbf{MP} & \textbf{LS} & \textbf{GW} & \textbf{KD} & \textbf{Acc}                \\
    \hline\hline
        &   &  &  &  &  92.46 (baseline)                      \\
       \checkmark &  &  &  &  & 95.63                              \\
        & \checkmark &  &  &  &  93.64                               \\
        &  & \checkmark &  &  &  92.98                                \\
        &  &  & \checkmark &  &  93.26                                 \\
        &  &  & & \checkmark  &  93.66                                  \\
        \hline
        \checkmark  & \checkmark  & & & & 96.25                         \\
        \checkmark & \checkmark & \checkmark & & & 96.50                \\
        \checkmark & \checkmark & \checkmark & \checkmark & & 96.98      \\
        \checkmark & \checkmark & \checkmark & \checkmark & \checkmark & \textbf{97.19} \\
    \hline
  \end{tabular}
\end{table}

\section{Conclusions}

In this paper, we introduce a convolutional neural networks based model for acoustic pornography recognition. We also research a series of refinements to improve model accuracy. General refinements including warmup, label smoothing and knowledge distillation introduce minor modifications to learning rate schedule or loss function and improve model performances consistently. Targeted refinements for acoustic pornography recognition such as coordinate attention and max pooling are powerful to detect pornographic audios according to the characteristics of them. Comprehensive experiments on our newly-collected large dataset demonstrate the effectiveness of our method. Specifically, the baseline model achieves 92.46\% accuracy and when all the refinements are stacked together with  appropriate settings, the accuracy is further improved to 97.19\%.  It is hoped that these model design and refinements will be of reference to the future development of acoustic pornography recognition. 

\bibliographystyle{IEEEbib}
\newpage
\bibliography{mybib}
\end{document}